\documentclass[usenatbib]{mn2e}
\usepackage{psfig}

\title{SSSPM~J0829$-$1309: A new nearby L dwarf detected in SuperCOSMOS 
Sky Surveys}
\author[R.-D. Scholz \& H. Meusinger]
       {R.-D. Scholz,$^{1,3}$\thanks{E-mail: rdscholz@aip.de} 
        H. Meusinger$^{2,3}$ \\
        $^1$Astrophysikalisches Institut Potsdam, 
              An der Sternwarte 16, D--14482 Potsdam, Germany\\
        $^2$Th\"uringer Landessternwarte Tautenburg, Sternwarte 5,
              D--07778 Tautenburg, Germany\\
        $^3$Visiting astronomer, German-Spanish Astronomical Centre, 
            Calar Alto, operated by the Max-Planck-Institute for Astronomy, \\
            Heidelberg, jointly with the Spanish National Comission for
            Astronomy}

\date{Accepted ...
      Received ...;
      in original form ...}

\pagerange{\pageref{firstpage}--\pageref{lastpage}}
\pubyear{2002}

\begin{document}

\maketitle

\label{firstpage}

\begin{abstract}

The SuperCOSMOS Sky Surveys provide a complete coverage of the Southern sky
in three passbands (photographic $B_J, R$ and $I$) and at different epochs
\citep{hambly01a,hambly01b,hambly01c}. These data are the basis for a new
high proper motion survey which aims at finding extremely red nearby dwarf 
stars and brown dwarfs. One of the first candidates, which is relatively 
bright ($I=16$) but very red ($R-I=2.8, B_J-R=3.6$), was detected
in the equatorial zone by its large proper motion of 0.56~arcsec/yr. 
Spectroscopic follow-up observations with the 2.2m telescope of the
Calar Alto Observatory classified this object as L2 dwarf very similar to
the first free-floating L dwarf Kelu~1 also discovered in a proper motion
survey by \citet{ruiz97}.
If we assume SSSPM~J0829$-$1309 to have the same luminosity as Kelu~1, we 
get a distance estimate for the new L dwarf of about 12~pc since it is about
one magnitude brighter than Kelu~1 in the SSS $I$ and $R$ bands. This makes
SSSPM~J0829$-$1309 one of the nearest objects of its class, well suited for
detailed investigations. We present a brief
overview of all known nearby ($d<20$~pc) southern L dwarfs and give first
proper motion values for DENIS-P~J0255$-$47 and SDSSp~J1326$-$00 and
an improved proper motion for LHS~102B.

\end{abstract}


\section{Introduction}

Only five years ago, the first field L dwarfs were discovered in a proper 
motion survey using ESO Schmidt plates \citep{ruiz97} and in the 
DEep Near-Infrared Survey (DENIS) 
\citep{delfosse97}. In the following years, the new spectral type ''L'' 
was defined using the rapidly growing number of discoveries from
the 2-Micron All-Sky Survey (2MASS) \citep{kirkpatrick99} and from DENIS
\citep{martin99}. Most of the additional L dwarfs found since then are
from 2MASS \citep{kirkpatrick00,reid00,gizis00} and from the Sloan
Digital Sky Survey (SDSS) \citep{fan00,schneider02,hawley02}.

The colour-based search for L dwarfs with DENIS, 2MASS and SDSS data has
increased the sample size to more than 200, where the majority of the objects 
lie at distances larger than 25~pc, the limit of the Nearby Star Catalogue
\citep{gliese91}.  Nearby L dwarfs, suitable for detailed follow-up 
observations, have also been found (see the list of \citealt{kirkpatrick00} 
and references therein). However, many nearby L stars are still waiting
to be discovered according to the predictions on their number densities
from observations \citep{reid99,gizis00} and theory \citep{chabrier02}.
The number density of early L dwarfs (L0.0-L4.5) alone is expected to be
around 0.002 per cubic parsec \citep{gizis00}, corresponding to about 
70 objects within 20~pc.
The completion of the census of L dwarfs in the Solar neighbourhood is
important for investigations of the star formation process, the stellar 
and substellar luminosity function and the initial mass function.
 
As demonstrated with the recent discovery of three nearby L dwarfs in the
Southern sky \citep{lodieu02}, the SuperCOSMOS Sky Surveys (SSS) 
\citep{hambly01a,hambly01b,hambly01c} provide an effective tool to find
nearby extremely red and low-luminosity objects via their proper motion
and photographic colour. The SSS cover the whole Southern sky with scans of 
UK Schmidt Telescope (UKST) plates in three different passbands ($B_J, R, I$) 
and at different epochs with additional scans of ESO Schmidt plates. Since
practically all very nearby stars ($d<10$~pc) in the 
Nearby Star Catalogue \citep{gliese91} 
have large proper motions ($\mu>0.18$~arcsec/yr, i.e. the
limit of the New Luyten Two Tenths (NLTT) catalogue, \citealt{luyten7980}), we
can expect large proper motions for all nearby L dwarfs as well. The reason
for the non-detection of L dwarfs among the NLTT stars is the lower
limiting magnitude of the first Palomar Observatory Sky Survey (POSS-1) 
used by Luyten, compared to later
Schmidt telescope surveys (POSS-2, UKST, ESO). UKST and ESO Schmidt plates
are much deeper with limiting magnitudes of
$B_J\sim23, R\sim22, I\sim19$ \citep{hambly01a}. It was thus possible to
find the first free-floating L dwarf in the Solar neighbourhood, Kelu~1 with
$R=19.7, I=17.1$ (SSS magnitudes) as high proper motion object
on Schmidt plates \citep{ruiz97}. Trigonometric parallax
measurements \citep{dahn02} have shown this benchmark L2 dwarf to lie at 
a distance of 19~pc. 

In this paper, we announce the detection of another nearby L dwarf
using the proper motion search technique.

\section{Finding nearby L dwarfs in SSS}

In order to estimate the chances to find nearby L dwarfs using SSS data, we 
have looked through the list of known objects ($d<20$~pc) in the southern sky.
Surprisingly, all known L dwarfs could be identified at least in the $I$
band data of the SSS, except one object, for which no $I$ plate measurement
was available (Table~\ref{sssdata}). 
This does not mean that we are able to find all
these objects in a proper motion survey based on SSS, since we need 
measurements at several epochs for that task. Since all early L dwarfs ($<$L5)
with $d<20$~pc were detected in both $R$ and $I$ data, there is a good chance
to discover this class of objects as proper motion objects if the epoch 
difference is not too
small. The epoch differences between $R$ and $B_J$ plates are typically
10-15 years. Those between $R$ and $I$ plates are smaller but with at least
a few years in many fields still useful. Even for late L dwarfs 
within 20~pc appearing only in the $I$ band data, there is the chance to
find them in the overlap area of the Schmidt plates of different epochs.
Former attempts to detect L dwarfs in a proper motion survey based on the
APM measurements of $B_J$ and $R$ plates \citep{scholz00} did not lead to
success since only early L dwarfs may show up in the $B_J$ band and the 
search area was restricted to a few thousand square degrees.

There are different possibilities to use the SSS data in a search for 
extremely
red proper motion objects. In \citet{lodieu02}, the basic search strategy 
started with looking for objects in the SSS catalogues
which were not matched with a corresponding
object in a different passband to within a nominal search
radius of 3~arcsec. Here we used the opposite strategy: in order to find 
nearby brown
dwarf candidates, we set the conditions that 1) an object was measured in all 
three passbands within a matching radius of 6~arcsec, 2) the SSS computed proper
motion (obtained from comparison of $B_J$ and $R$ plates) is larger than
100~mas/yr, 3) the object has star-like image parameters and 4) the colours are 
extremely red, i.e. $B_J-R>3$ and $R-I>2$. The proper motion determination
was then improved by further positional information from the $I$ plates and
from 2MASS (if available). Only the equatorial zone 
($-17^{\circ} < \delta < +3^{\circ}$) suitable for follow-up observations
from Calar Alto was included in the search. The target of the present paper 
was selected as one of the most promising candidates in that search. It is
called SSSPM~J0829$-$1309 since it was found via its proper motion in SSS data
(see Table~\ref{sssdata}). 

\begin{table*}
\begin{minipage}{175mm}
 \caption[]{SSS proper motions and photometry of SSSPM~J0829$-$1309 
compared to all known nearby ($d<20$~pc), southern
L dwarfs.} 
\label{sssdata}
 \begin{tabular}{rlcccrccccr}
 \hline
No. &Name & $\alpha, \delta$ & Epoch & $\mu_{\alpha}\cos{\delta}$ & 
$\mu_{\delta}$ & $B_J$ & $R$ & $I$ & Sp. & $d$ \\
    & & (J2000) & & \multispan{2}{\hfil mas/yr \hfil} & SSS & SSS & SSS &  & 
pc \\
 \hline
 1 & SSSPM~J0829$-$1309&08 28 34.11 $-$13 09 20.1&2001.28& $-565\pm13$&   
$+4\pm21$ & 22.58 & 18.83 & 16.01 & L2 & 12  \\
 2 & SSSPM~J0109$-$5101&01 09 01.29 $-$51 00 51.1&1990.80& $+207\pm04$&  
$+94\pm11$ & 21.19 & 18.21 & 14.81 & L2 & 13  \\
 3 & Kelu~1            &13 05 40.18 $-$25 41 06.0&1998.41& $-289\pm14$&  
$-24\pm13$ &       & 19.73 & 17.11 & L2 & 19  \\
 4 &DENIS-P~J1058$-$15&10 58 47.84 $-$15 48 17.5&1998.23& $-287\pm73$&  
$+63\pm36$ &       & 20.30 & 17.63 & L3 & 17  \\
 5 & 2MASSW~J2224$-$01 &22 24 43.82 $-$01 58 52.5&1998.77& $+533\pm38$& 
$-910\pm31$ &       & 20.65 & 18.40 & L4.5 & 11  \\
 6 & LHS~102B          &00 04 34.66 $-$40 44 01.6&1996.87& 
$+648\pm13$&$-1476\pm05$ &       & 19.61 & 16.88 & L5   & 10  \\
 7 & SDSSp~J0539$-$00  &05 39 51.99 $-$00 59 02.0&1998.83&            &       
      &       &       &       & L5   & 14  \\
 8 & DENIS-P~J1228$-$15&12 28 15.23 $-$15 47 34.2&1998.25& $+133\pm58$& 
$-242\pm01$ &       & 20.59 & 17.72 & L5   & 20  \\
 9 & 2MASSW~J1507$-$16 &15 07 47.70 $-$16 27 38.5&1998.34& $-137\pm17$& 
$-908\pm01$ &       & 19.23 & 16.14 & L5   &  7  \\
10 & DENIS-P~J0205$-$11&02 05 29.35 $-$11 59 30.4&1998.60& $+473\pm02$&  
$+38\pm125$&       &       & 17.93 & L7   & 20  \\
11 & DENIS-P~J0255$-$47&02 55 03.59 $-$47 00 50.9&1998.93& $+997\pm28$& 
$-591\pm16$ &       & 20.10 & 17.00 & L8   &  5  \\
12 & SDSSp~J1326$-$00  &13 26 29.83 $-$00 38 31.2&1999.09& 
$-268\pm...$&$-1004\pm...$ &       &       & 18.33 & $\sim$L8 & 18 \\
 \hline
 \end{tabular}
 \medskip
{\bf Notes:}

Positions are given for the latest available epoch (in most cases 
the 2MASS positions). SSS magnitudes are average values in case of overlapping 
plates of the same passband. The discovery paper as well as the reference for 
the distance estimate (rounded to 1~pc, but uncertainties are smaller for all 
objects with trigonometric parallaxes measured by \citealt{dahn02}) and some 
details on the proper motion determination are given in the notes below 
(corresponding to No. in first table column). 
Spectral types are in the system of
\citet{kirkpatrick99}. \citet{martin99} adopted different spectral types
for DENIS-P~J0255$-$47 (L6), DENIS-P~J0205$-$11 (L5), DENIS-P~J1228$-$15
(L4.5), LHS~102B (L4) and DENIS-P~J1058$-$15 (L2.5).

{\bf 1} -- This paper. The proper motion is based on six positions, as 
measured on two $B_J$, two $R$ and two $I$ plates within the SSS. Distance
estimate is based on the comparison of photographic SSS magnitudes
with those of Kelu~1. 

{\bf 2} -- \citet{lodieu02}. Proper motion was obtained from four SSS 
positions (one $B_J$, two $R$ and one $I$ plates). The distance estimate was
based on the comparison of the infrared $K_s$ magnitude measured by 
\citet{lodieu02} with that of Kelu~1. Note that the spectral type was 
determined from infrared observations only with an uncertainty of one 
spectral 
class \citep{lodieu02}. If we estimate the distance by comparison of the 
measured SSS magnitudes with those of Kelu~1, we get an even smaller distance 
of $\sim8$~pc. 

{\bf 3} -- \citet{ruiz97}. The proper motion is obtained from four SSS 
positions (two $R$ and two $I$ plates) and the 2MASS position. 
The total proper motion is smaller than given in 
\citet{ruiz97} but in good agreement with the more accurate value given by 
\citet{dahn02}, where the distance is also taken from. 

{\bf 4} -- \citet{delfosse97}. Proper motion is 
obtained from two SSS positions combined with the 2MASS position. Due to the 
small time baseline of only 6 years, the proper motion accuracy is low. 
However, the proper motion is in good agreement with the high accurate value
determined by \citet{dahn02}. Distance is from \citet{dahn02}. 

{\bf 5} -- \citet{kirkpatrick00}. Proper motion is from four SSS (two $R$ and 
two $I$ plates) and the 2MASS positions. It is in reasonable good agreement
with the accurate proper motion determined by \citet{dahn02}. Distance is 
from 
\citet{dahn02}.

{\bf 6} -- \citet{goldman99}. Proper motion was obtained from 6 SSS positions 
(two $B_J$, two $R$ and two $I$ plates). The proper motion of the primary,
LHS102A has also been obtained (from 8 SSS positions including two $B_J$
plates), and is slightly different from that of LHS102B: $+678\pm03$ and
$-1506\pm03$~mas/yr. The small (30~mas/yr) differences in both directions
may be due to orbital motion. Unfortunately, LHS102A with $B_J=14.19, R=11.88$
and $I=9.85$ was too faint to be measured in the AC2000 catalogue 
\citep{urban98} so that no accurate astrometric measurements at epochs before
1977 (the SSS $B_J$ plates) are available. Distance (of LHS102A) is from 
\citet{vanaltena95}.

{\bf 7} -- \citet{fan00}. Not detected in SSS $B_J$ and $R$ measurements.
SSS $I$ measurements were not yet available. Distance estimate is from
\citet{kirkpatrick00}.

{\bf 8} -- \citet{delfosse97}. Position is from 2MASS. The position given in
the discovery paper does not agree with the finding chart \citep{delfosse97}.
The proper motion given in the table has been obtained from two SSS positions
and the 2MASS position. Despite its large errors, it is in reasonable good
agreement with the accurate value of \citet{dahn02}. Distance is 
from \citet{dahn02}. 

{\bf 9} -- \citet{reid00}. Proper motion was obtained
from the 2MASS and two SSS positions and is in good
agreement with the accurate value of \citet{dahn02}.
Distance is from \citet{dahn02}. 

{\bf 10} -- \citet{delfosse97}. Proper motion is from 2MASS  
combined with two SSS $I$ plates measurements. Despite the large 
proper motion errors, there is good agreement with the accurate proper motion
of \citet{dahn02}. Distance is from \citet{dahn02}. 

{\bf 11} -- \citet{martin99}. The proper motion is obtained
from four SSS positions (two $R$ and two $I$ plates) and the 2MASS postion.
No previous proper motion determination was known. Distance estimate is from 
\citet{kirkpatrick00}.

{\bf 12} -- \citet{fan00}. Not detected in SSS $B_J$ and $R$ measurements.
Proper motion is from comparing only two positions: SSS $I$ and 2MASS,
therefore, no proper motion errors are given.
Distance estimate is from \citet{kirkpatrick00}. 

\end{minipage}
\end{table*}

All measured proper motions of the objects in Table~\ref{sssdata} are larger
than the NLTT limit of 0.18~arcsec/yr.
Proper motions obtained by \citet{dahn02} are much more accurate than the 
SSS based proper motions in Table~\ref{sssdata}. However, for objects
not included in \citet{dahn02} the use of SSS data allowed a first proper
motion determination (DENIS-P~J0255$-$47, SDSSp~J1326$-$00) or led to an 
improved proper motion (LHS~102B).

\begin{figure*}
\begin{minipage}{120mm}
 \psfig{figure=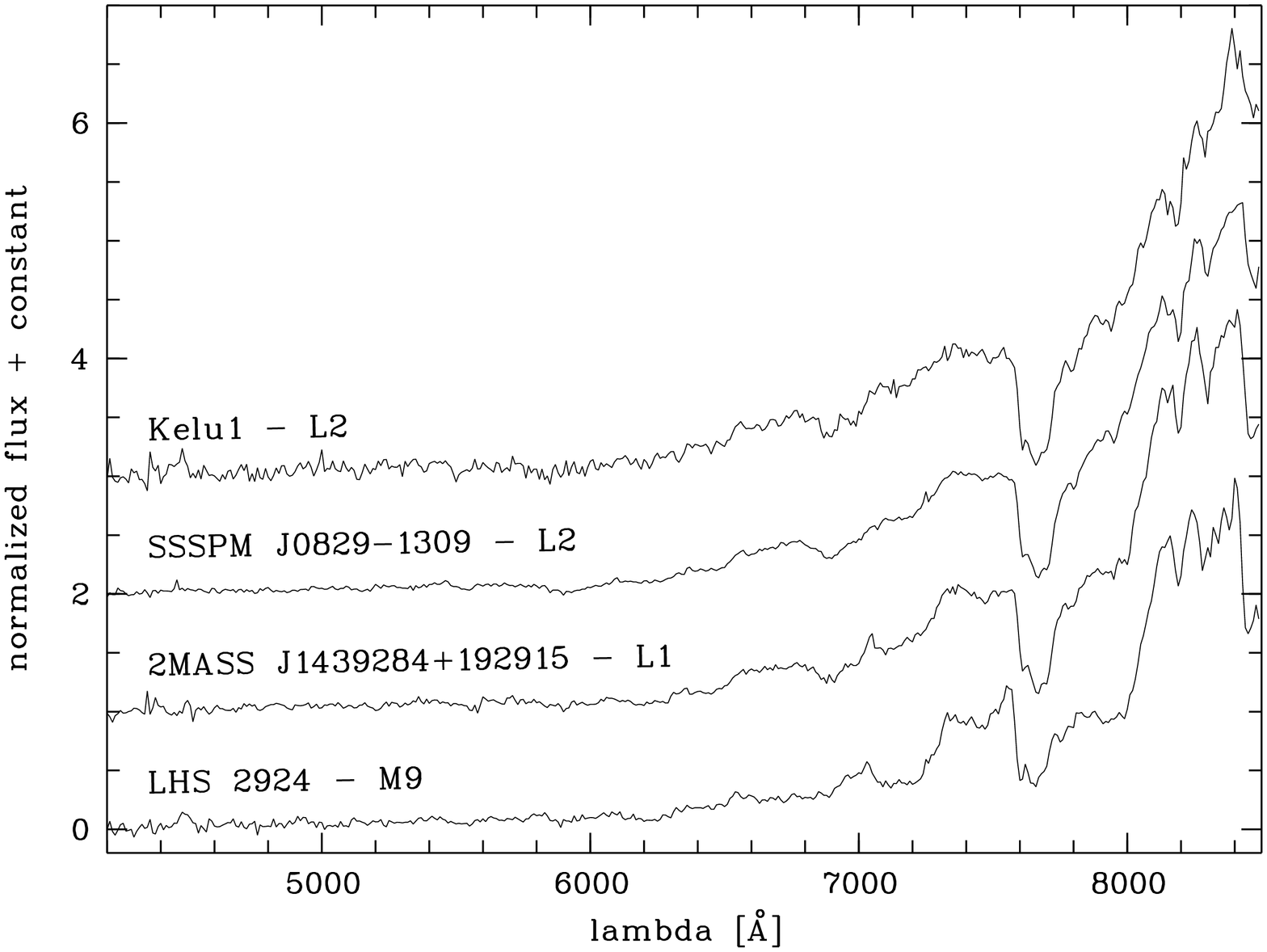,width=120mm,angle=270,clip=}
 \caption{Spectrum of SSSPM~J0829$-$1309 in comparison to those of other 
late-M and early-L dwarfs. All spectra were obtained with 
CAFOS at the 2.2m Calar Alto telescope ($F_{\lambda}$ normalised at
7500\AA). Spectral types of the comparison
objects are from \citet{kirkpatrick00} and \citet{reid00}. Note the 
disappearance of the TiO band (7053\AA\, head, degraded to red) and
the VO bands (broad trough $\sim$7334--7534\AA\, and $\sim$7851--7973\AA)
from spectral type M9 to L2 \citep{kirkpatrick99}.}
 \label{spec4}
\end{minipage}
\end{figure*}

\section{Spectroscopic distance estimate}

Spectroscopic follow-up observations were carried out with CAFOS, the
focal reducer and faint objects spectrograph at the 2.2m Calar Alto telescope.
The grism B-400 was used with a wavelength coverage from
4000\AA\, to 10000\AA. The combination with the SITe-1d CCD
(0.53~arcsec per pixel)  yields 10\AA\, per pixel. A slit width of
1~arcsec was used. The exposure time
was 45 minutes both for SSSPM~J0829$-$1309 and for Kelu~1, 23 minutes for
2MASS~J1439284$+$192915 and 6 minutes for LHS~2924. 
The spectra of
SSSPM~J0829$-$1309 and the comparison objects were taken under very 
similar atmospheric conditions and at comparable airmass. Wavelength and flux
calibration was done under MIDAS using the Calar Alto standard reduction 
procedure SPECTRUM (written by K. Meisenheimer).
The wavelength calibration was checked by means of night sky lines.  

The resulting spectra of SSSPM~J0829$-$1309 and
three comparison objects are shown in Figure~\ref{spec4}. 
We classify SSSPM~J0829$-$1309 as an L2 dwarf
due to the similarity of its spectrum with that of Kelu~1 (decreasing 
strength of TiO and VO absorption), 
which can also be seen if compared to the higher 
signal-to-noise spectra of Kelu~1 shown in \citet{ruiz97} and \cite{reid00}. 

There are no 2MASS magnitudes available for SSSPM~J0829$-$1309.
The distance estimate has therefore been obtained from a comparison 
with the photographic SSS magnitudes of Kelu~1, for which a trigonometric
parallax ($d=18.7$~pc) has been determined by \citet{dahn02}. Our target is 
roughly one magnitude brighter in both $R$ and $I$ so that we get a distance 
estimate of approximately 12~pc assuming the same absolute magnitude as for
Kelu~1.

We have also obtained a higher resolution spectrum of SSSPM~J0829$-$1309
using grism R-100 (2\AA\, per pixel) with 1 hour exposure time. 
Low signal-to-noise features in this spectrum are consistent with
Lithium absorption (at 6708\AA) and $H_{\alpha}$ emission. This 
interpretation, if true, would make SSSPM~J0829$-$1309 a brown dwarf similar 
to Kelu~1, but a confirmation by a higher signal-to-noise spectrum 
is needed.

\section{Conclusions}

We have discovered one of the nearest field L dwarfs, SSSPM~J0829$-$1309, 
which we classify as L2 dwarf similar to Kelu~1. Based on the comparison 
of SSS magnitudes of these two objects we estimate SSSPM~J0829$-$1309 to
lie at a distance of about 12~pc.
The newly discovered L dwarf is a potential target for both Northern and 
Southern parallax programmes.

\section*{Acknowledgements}
This research has made use of data products from the SuperCOSMOS Sky Surveys
at the Wide-Field Astronomy Unit of the Institute for Astronomy, University
of Edinburgh. The authors would like to thank Nigel Hambly for advise on the 
use of SSS data.

The spectroscopic classification and distance estimate is based on 
observations made  with the 2.2\,m telescope of the German-Spanish 
Astronomical Centre, Calar Alto, Spain. We would like to thank the staff 
of the Calar Alto observatory for their help during the observations.


\bsp 

\label{lastpage}

\end{document}